\documentclass[12pt]{article}

\usepackage{amsmath}
\usepackage{amssymb}
\usepackage{amsthm}
\usepackage[utf8]{inputenc}
\usepackage{graphicx}

\newcommand*{\E}{\mathbb{E}}

\usepackage[]{algorithm2e}

\renewcommand{\and}{\text{ and }}

\allowdisplaybreaks

\theoremstyle{definition}

\allowdisplaybreaks
\theoremstyle{remark}

\numberwithin{equation}{section}
\title{Scaling Properties of Weakly Self-Avoiding Fractional Brownian Motion in One Dimension}
\author{{\bf Wolfgang Bock}\\
Technomathematics Group, TU Kaiserslautern, Germany \\
bock@mathematik.uni-kl.de\\[.3cm]
{\bf Jinky B.~Bornales}\\
MSU-IIT, Iligan, Philippines\\
jbornales@gmail.com\\[.3cm]
{\bf Cresente O.~Cabahug}\\
MSU-IIT, Iligan, Philippines\\
cresentekudo@gmail.com\\[.3cm]
{\bf Samuel Eleutério}\\
CFTP-IST, Univ. Lisbon, Portugal\\
sme@ist.utl.pt\\[.3cm]
{\bf Ludwig Streit}\\
BiBoS, Univ. Bielefeld, Germany \\
and CCM, Univ. Madeira, Portugal\\
streit@uma.pt}

\begin{document}
\maketitle
\abstract{We use an off-lattice discretization of fractional Brownian motion and a Metropolis Algorithm to determine the asymptotic scaling of this discretized fractional Brownian motion under the influence of an excluded volume as in the Edwards and Domb-Joyce models. We find a good agreement between the Flory index describing the scaling of end-to-end length with a mean field formula proposed earlier for this class of models.}

\section{Introduction}
\subsection{Background}
Random paths and in particular (weakly) self-avoiding paths have been studied intensively in physics as well as in probability theory. In physics they play  a crucial role in the modeling of chain polymers with "excluded volume effect".  In these models self-intersections are penalized; in the continuum case this is the Edwards model \cite{Edwards}, in the discrete case one speaks of the Domb-Joyce model \cite{Domb-Joyce}.
The penalization factor in the Edwards model is introduced through a modification of the Wiener measure $\mu_0$ by a Gibbs factor 
$$
d \mu = \frac{1}{Z} \exp\left(-g \int_0^N \, d\tau \int_0^\tau \, dt \, \delta(B(\tau)-B(t))\right) \, d\mu_0,
$$
where $N,g>0$, $\delta$ is the $d$-dimensional Dirac delta function, $B$ denotes a version of a $d$-dimensional Brownian motion
and
$$
Z= \E\left(\exp\left(-g \int_0^N \, d\tau \int_0^\tau \, dt \, \delta(B(\tau)-B(t))\right)\right).
$$
We set
$$
L\equiv  \int_0^N \, d\tau \int_0^N \, dt \, \delta(B(\tau)-B(t)),
$$
which is known as the "self-intersection local time" of a Brownian motion.

The one-dimensional weakly self-avoiding random walk is well understood and proofs of mathematical rigor concerning the scaling exponent can be found in \cite{greven1993}. 

For fractional weakly self-avoiding paths the contrary is the case. This is due to the lack of Markov and martingale properties, which excludes many techniques of stochastic analysis. \\

\subsection{Fractional Model}
Fractional Brownian motion (fBm) has been suggested as a more general model for chain polymers, see e.g.~\cite{Hartmann13}.\\
FBm on $\mathbb{R}^{d}$, $d\geq 1$, with "Hurst
parameter" $H\in \left( 0,1\right) $ is a $d$-tuple of independent centered Gaussian processes \cite{oks,Mishura} $B^{H}=\{B_{t}^{H}:t\geq 0\}$ with
covariance function 
\begin{equation}\label{BH} 
\mathbb{E}(B_{t}^{H}B_{\tau}^{H}) =\frac{1}{2}\left(
t^{2H}+\tau^{2H}-|t-\tau|^{2H}\right).
\end{equation}

One sees that, for $H=1/2$, one recovers ordinary $d$-dimensional Brownian motion. \
From (\ref{BH}) one further verifies stationary increments:%
$$
\mathbb{E}\left( \left( B^{H}(t)-B^{H}(\tau)\right) ^{2}\right) =\left\vert
t-\tau\right\vert ^{2H},
$$
and since the increments are centered Gaussians, we have more generally%
$$
\mathbb{E}\left( \left( B^{H}(t)-B^{H}(\tau)\right) ^{2k}\right) =c_{k}\left\vert
t-\tau\right\vert ^{2Hk}, \quad c_k >0,
$$
which, by the Kolmogorov-Chentsov theorem implies the continuity of sample
paths. Increments are uncorrelated for $H=1/2$, correlated ("persistent" paths, or
stiffer polymers) for $H>1/2$, and anti-correlated \ (curlier than Brownian
motion) for small Hurst indices $H<1/2.$  Because of these properties fractional Brownian motion paths have been proposed as generalized chain polymer models by \cite{BC95}
A rigorous mathematical extension of the Edwards model was recently shown
\cite{GOSS} to exist for a limited range of Hurst parameters $H$ and
dimensions $d.$ This limitation is due to the singular nature of the
self-intersection local time and will not be shared by discretizations which
are thought to be in the same universality class of asymptotic scalings.
The characteristic observable of this general class of chain polymer models is the average mean-squared end-to-end length $R_e^2$ and its asymptotic scaling behavior as the number of monomers increases. 
In this paper we intend a first exploration of the scaling behavior for fBm-based models. We should emphasize that we do not yet aim for high precision results in this first study.
In this exploratory phase we have restricted ourselves to the one-dimensional case. There the mean end-to-end length of paths grows linearly for self-avoiding random walks \cite{vanderhofstad1997}, as is to be expected intuitively. For fractional paths with $H< \frac{1}{2}$ on the other hand the situation is far from being obvious: the repulsive excluded-volume dynamics is balanced by long-range attractive force acting along fractional Brownian trails for small Hurst index. These models might be considered unphysical. However, as in the Brownian case, there exists a recursion formula \cite{KosmasFreed78}, which would predict the Flory index in higher dimensional cases from the one-dimensional one, \cite{BOS11}.
\section{The Algorithm}
\subsection{Discretization and Metropolis Algorithm}
There are various methods to simulate fBm paths. Exact methods such as the ones of Hosking \cite{Hosking}, the Cholesky method, see e.g.~\cite{Asmussen} or the method of Wood and Chan \cite{WoodChan}, which uses roots of the covariance to obtain a fractional path from a standard normal distribution are not usable here, since it is not obvious how to include directly the excluded volume. The same holds for approximative methods using spectral densities, the Paxson method \cite{Paxson}, which uses Fourier transform or the Decreusefond-Lavaud method \cite{Decreusefond} which is using the kernel representation of fBm. \\
A more natural way to implement a discrete fractional walk with self-repellence is to use off-lattice discretization and Monte Carlo methods based on a Metropolis algorithm as in \cite{Besold} for the Brownian case and in \cite{Hartmann13} for the fractional Brownian motion. 
In the our algorithm the Metropolis routine is used to update "bonds", i.e.~increments of the path.
Given a Hurst parameter $H\in (0,1),$ we
define the positions of  a walk with $N$ points by %
\[
x_{j}=B^{H}(j),\quad j=0,\ldots ,N-1,
\]
and $N-1$ "bond vectors" 
\[
y_{j}=B^{H}(j+1)-B^{H}(j).
\]%
These are centered Gaussian, so one obtains their probability density by
inverting the covariance $A_{jl}=\mathbb{E(}y_{j}y_{l})$. 
\[
\varrho _{0}(y)=\frac{1}{\sqrt{\left( 2\pi \right) ^{N-1}\det A}}\exp \left(
-\frac{1}{2}(y,A^{-1}y)\right) \equiv C\exp \left( -\frac{1}{2}%
(y,\mathcal{H}_{0}y)\right) .
\]
To obtain a one-dimensional weakly self-avoiding fractional random walk we have to penalize self-crossings. To
discretize the self-intersection local time: 
\[
L=\int_{0}^{N}\,d\tau\,\int_{0}^{N}\,dt\, \delta \left( B^{H}(\tau)-B^{H}(t)\right)
=\int_{\mathbb{R}}\,d u\,L_{u}^{2},
\]%
with 
\[
L_{u}=\int_{0}^{N}\,dt\, \delta \left( B^{H}(t)-u\right) ,\quad u\in \mathbb{R%
},
\]%
we decompose $\mathbb{R}$ into intervals $I_{n}$ of equal length $l$ and
replace $L_{u}$ by the number of positions $x_{k}$ that fall into the interval
with number $n$:\newline
\[
L_{n}=\# \{x_{j}|x_{j}\in I_{n}\}.%
\]
Likewise we replace the self-intersection local time by 
$$
L=\sum_{n} L_{n}^{2},
$$
hence the unnormalized probability density of conformations becomes 
\[
\rho (x)\sim \exp \left( -\frac{1}{2}(y,\mathcal{H}_{0}y)-g(L(y))\right) .
\]%
The well-known Metropolis algorithm, see e.g.~\cite{Asmussen} in our case with the energy 
\begin{equation}\label{energytotal}
E=(y,\mathcal{H}_{0}y)-gL(y),
\end{equation}
does not require normalization of the probability density. In the above formula $y\in \mathbb{R}^{N-1}$ is the vector of increments and $g\geq 0$ is
the coupling constant of the excluded-volume term $L$.
\subsection{Qualitative Description of the Algorithm}
For a fixed polymer length $N$ we start from a random initial configuration, randomly chosen bond vectors are updated and subjected to the standard Metropolis routine.
After an initial relaxation phase $Nr$, the end-to-end lengths are sampled, not after individual updates but in intervals large enough to suppress correlations.  Finally the average of the sample is determined. The process then is repeated for increasing $N$ to determine the scaling of the end-to-end length with growing $N$.\\
The error analysis was based on standard analysis of $30$ independent runs with in particular individual random seeds.

\subsection{Parameter Choice}

For the number of monomers $N$ our choice of the range between $200$ and $600$ was dictated by two considerations. $N$ should be large enough to approximate the asymptotic regime. The upper limit was chosen such that equilibrium could be reached within a few days of computing time.  
The interval length $l$ in the definition of incidences above was chosen to be $l=1$.

\begin{figure}[h]\label{figure_overshoot}
\includegraphics[scale=.5]{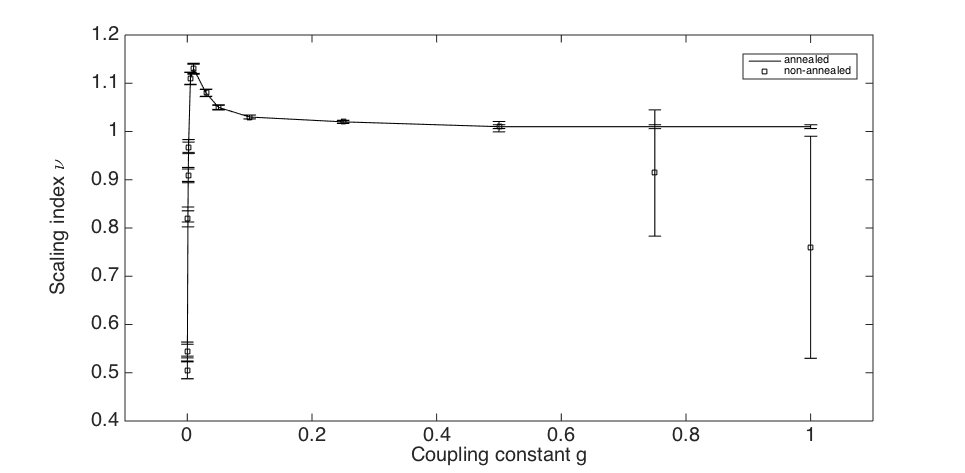}
\caption{Dependence of the scaling exponent from the penalizing strength $g$ (for $r=3$ and $H=0.5$, $N=100, 120, 160$ and $200$, $s=25 \times 10^6$ updates and $Nr=5\times 10^6$ relaxation time with $30$ repetitions).}
\end{figure}

Recall that $g$ is the coupling constant or strength of excluded volume effect. For $g=0$ the walks have no penalization and will scale as a fBm-path, i.e.~the scaling exponent will be $\nu=H$. For small positive $g$ the scaling index $\nu$ rises sharply, see Figure 1 for the Brownian case. It overshoots the theoretical value and then seemingly drops below it for larger $g$. Scrutiny of the end-to-end length in this regime reveals the cause of this. As $g$ increases there appears an increasing number of "outliers" with much shorter end-to-end length, i.e. conformations that did not unfold to equilibrium conformations during the relaxation and sampling period. To amend this problem we did not perform relaxations with a constant $g$, instead we slowly increased to coupling constant from zero, with the result that the outliers disappeared and the scaling index stabilized. See the black line in Figure 1.
\begin{figure}[h]\label{figure_rtest}
\centering
\includegraphics[scale=.5]{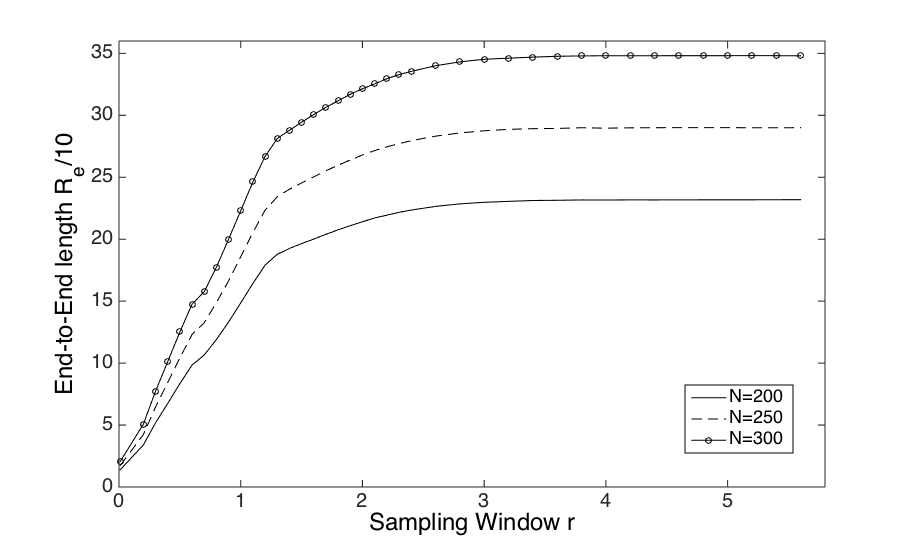}
\caption{End-to-end length as a function of sampling window $r$ (for $g=2$,$H=0.5$, $N=200, 250, 300$,  $s=25 \times 10^6$ and $Nr=5\times 10^6$ with $30$ repetitions).}
\end{figure}
Monte Carlo updates of bond vectors are chosen from a sampling window $[-r,r]$. As a function of $r$ the computed end-to-end length will saturate within the sampling period when $r$ is large enough, see Figure 2. This led us to choose $r=4$. Larger sampling windows would increase the rate of rejects in the Metropolis algorithm. \\

An analogous analysis for Hurst parameters $H<\frac{1}{2}$ produces the same results for suitable $g$ and $r$.\\

One observes that the relaxation of end-to end length is some orders of magnitudes slower than that of the two energies in equation \eqref{energytotal}, \cite{Cabahug14}.
After some studies we chose  $5\times 10^6$ relaxation updates $Nr$ as a matter of computational convenience; which proved to be sufficient when we modified the relaxation protocol as described above to facilitate unfolding. 
The number of updates in the sampling phase was $s=25\times 10^6$, and sampling was performed every $10000$ updates.

\section{Results and Conclusions}
\subsection{The Flory index}
While there are not yet any mathematical proofs it is generally expected that for large $N$ the end-to-end length of polymers will scale according to $R_e^2 \sim N^{2\nu}$ when they are modeled by (weakly) self-avoiding random paths.
Mean field arguments - notoriously unsound and yet successful - would suggest the following formula for the scaling index in the fractional case
\begin{equation}\label{eq_BOSnu}
\nu = \frac{2H+2}{d+2},
\end{equation}
with $N \gg 1$ generalizing the famous Flory formula \cite{Flory} for $H=\frac{1}{2}$. \\
In Figure 3 the prediction in \eqref{eq_BOSnu} is compared to our numerical findings. 

\begin{figure}[h]\label{figure3}
\centering
\includegraphics[width=1.2\textwidth]{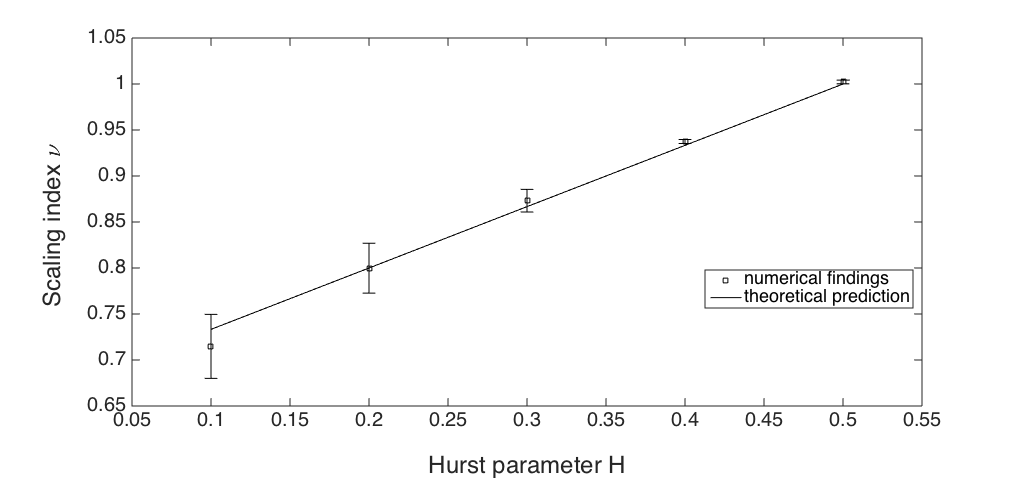}
\caption{Comparison of the prediction in \eqref{eq_BOSnu} and the numerical findings (for $g=3$,$r=4$, $H=0.5$, $N=200, 300, 400,500,600$,  $s=25 \times 10^6$ and $Nr=5\times 10^6$ with $30$ repetitions.)}
\end{figure}

$\;$\\
\noindent{\bf Acknowledgments:} 
 This work was financed by Portuguese
national funds through FCT - Funda\c{c}\~{a}o para a Ci\^{e}ncia e
Tecnologia within the project PTDC/MAT-STA/1284/2012.
The authors like to thank the IIT Iligan High Performance Computing for the opportunity to use their cluster for simulations. Moreover we like to thank Sim Bantayan and Roel Baybayon for discussions and remarks. We would like to thank DOST-ASTHRD, CHED-PHERNET and OVCRE (MSU-IIT) for the local support of Ludwig Streit during his research visits at MSU-IIT. We like to thank Maria João Oliveira for her helpful comments and support. 

\bibliographystyle{alpha}
\bibliography{numericbib}

\begin{thebibliography}{vdHdHK97}

\bibitem[AG07]{Asmussen}
S.~Asmussen and P.W. Glynn.
\newblock {\em Stochastic Simulation: Algorithms and Analysis}.
\newblock Applications of Mathematics. Springer, 2007.

\bibitem[BC95]{BC95}
P.~Biswas and B.J. Cherayil.
\newblock Dynamics of fractional brownian walks.
\newblock {\em The Journal of Physical Chemistry}, 99(2):816--821, 1995.

\bibitem[BGZ00]{Besold}
G.~Besold, H.~Guo, and M.~J. Zuckermann.
\newblock Off-lattice {M}onte {C}arlo simulation of the discrete {E}dwards
  model.
\newblock {\em Journal of Polymer Science Part B: Polymer Physics},
  38(8):1053--1068, 2000.

\bibitem[BH{\O}Z07]{oks}
F.~Biagini, Y.~Hu, B.~{\O}ksendal, and T.~Zhang.
\newblock {\em Stochastic Calculus for Fractional {B}rownian Motion and
  Applications}.
\newblock Probability and Its Applications. Springer, 2007.

\bibitem[BOS11]{BOS11}
J.~Bornales, M.J. Oliveira, and L.~Streit.
\newblock {\em Self-repelling fractional {B}rownian motion - a generalized
  {E}dwards model for chain polymers}, chapter~33, pages 389--401.
\newblock World Scientific, 2011.

\bibitem[Cab14]{Cabahug14}
C.O. Cabahug.
\newblock unpublished.
\newblock 2014.

\bibitem[CW97]{WoodChan}
G.~Chan and A.~Wood.
\newblock Algorithm as 312: An algorithm for simulating stationary {G}aussian
  random fields.
\newblock {\em Journal of the Royal Statistical Society. Series C (Applied
  Statistics)}, 46(1):pp. 171--181, 1997.

\bibitem[DJ72]{Domb-Joyce}
C.~Domb and G.~S. Joyce.
\newblock {Cluster expansion for a polymer chain}.
\newblock {\em Journal of Physics C: Solid State Physics}, 5(9):956+, 1972.

\bibitem[DL96]{Decreusefond}
L.~{Decreusefond} and N.~{Lavaud}.
\newblock Simulation of the fractional {B}rownian motion and application to the
  fluid queue.
\newblock In {\em Proceedings of the ATNAC'96 conference}, 1996.

\bibitem[Edw65]{Edwards}
S.~F. Edwards.
\newblock The statistical mechanics of polymers with excluded volume.
\newblock {\em Proc.~Phys.~Soc.}, 85:613--624, 1965.

\bibitem[Flo53]{Flory}
P.~J. Flory.
\newblock {\em Principles of Polymer Chemistry}.
\newblock Baker lectures 1948. Cornell University Press, 1953.

\bibitem[GdH93]{greven1993}
A.~Greven and F.~den Hollander.
\newblock A variational characterization of the speed of a one-dimensional
  self- repellent random walk.
\newblock {\em The Annals of Applied Probability}, 3(4):1067--1099, 11 1993.

\bibitem[GOdSS11]{GOSS}
M.~Grothaus, M.J. Oliveira, J.L. da~Silva, and L.~Streit.
\newblock Self-avoiding fractional {B}rownian motion—the {E}dwards model.
\newblock {\em Journal of Statistical Physics}, 145(6):1513--1523, 2011.

\bibitem[HMR13]{Hartmann13}
A.~Hartmann, S.~Majumdar, and A.~Rosso.
\newblock Sampling fractional {B}rownian motion in presence of absorption: A
  {M}arkov chain method.
\newblock {\em Phys Rev E Stat Nonlin Soft Matter Phys}, 88(2):022119, 2013.

\bibitem[Hos84]{Hosking}
J.~R.~M. Hosking.
\newblock Modeling persistence in hydrological time series using fractional
  differencing.
\newblock {\em Water Resources Research}, 20(12):1898--1908, 1984.

\bibitem[KF78]{KosmasFreed78}
M.~K. Kosmas and K.~Freed.
\newblock On scaling theories of polymer solutions.
\newblock {\em The Journal of Chemical Physics}, 69(8):3647--3659, 1978.

\bibitem[Mis08]{Mishura}
Y.~S. Mishura.
\newblock {\em Stochastic Calculus for Fractional {B}rownian Motion and Related
  Processes}.
\newblock Lecture Notes in Mathematics. Springer-Verlag Berlin Heidelberg,
  Berlin, Heidelberg, 2008.

\bibitem[Pax97]{Paxson}
V.~Paxson.
\newblock Fast, approximate synthesis of fractional {G}aussian noise for
  generating self-similar network traffic.
\newblock {\em Comput. Commun. Rev.}, 27(5):5--18, 1997.

\bibitem[vdHdHK97]{vanderhofstad1997}
R.~van~der Hofstad, F.~den Hollander, and W.~König.
\newblock Central limit theorem for the {E}dwards model.
\newblock {\em Ann. Probab.}, 25(2):573--597, 04 1997.

\end{thebibliography}
\end{document}